# TriDo-Former: A Triple-Domain Transformer for Direct PET Reconstruction from Low-Dose Sinograms

Jiaqi Cui[1], Pinxian Zeng[1], Xinyi Zeng[1], Peng Wang[1], Xi Wu[2], Jiliu Zhou[1,2], Yan Wang[1(✉)], and Dinggang Shen[3,4(✉)]

[1] School of Computer Science, Sichuan University, China
`wangyanscu@hotmail.com`
[2] School of Computer Science, Chengdu University of Information Technology, China
[3] School of Biomedical Engineering, ShanghaiTech University, China
[4] Department of Research and Development, Shanghai United Imaging Intelligence Co., Ltd., Shanghai, China
`dinggang.shen@gmail.com`

**Abstract.** To obtain high-quality positron emission tomography (PET) images while minimizing radiation exposure, various methods have been proposed for reconstructing standard-dose PET (SPET) images from low-dose PET (LPET) sinograms directly. However, current methods often neglect boundaries during sinogram-to-image reconstruction, resulting in high-frequency distortion in the frequency domain and diminished or fuzzy edges in the reconstructed images. Furthermore, the convolutional architectures, which are commonly used, lack the ability to model long-range non-local interactions, potentially leading to inaccurate representations of global structures. To alleviate these problems, in this paper, we propose a transformer-based model that unites triple domains of sinogram, image, and frequency for direct PET reconstruction, namely TriDo-Former. Specifically, the TriDo-Former consists of two cascaded networks, i.e., a sinogram enhancement transformer (SE-Former) for denoising the input LPET sinograms and a spatial-spectral reconstruction transformer (SSR-Former) for reconstructing SPET images from the denoised sinograms. Different from the vanilla transformer that splits an image into 2D patches, based specifically on the PET imaging mechanism, our SE-Former divides the sinogram into 1D projection view angles to maintain its inner-structure while denoising, preventing the noise in the sinogram from prorogating into the image domain. Moreover, to mitigate high-frequency distortion and improve reconstruction details, we integrate global frequency parsers (GFPs) into SSR-Former. The GFP serves as a learnable frequency filter that globally adjusts the frequency components in the frequency domain, enforcing the network to restore high-frequency details resembling real SPET images. Validations on a clinical dataset demonstrate that our TriDo-Former outperforms the state-of-the-art methods qualitatively and quantitatively.

# 1 Introduction

As an in vivo nuclear medical imaging technique, positron emission tomography (PET) enables the visualization and quantification of molecular-level activity and has been extensively applied in hospitals for disease diagnosis and intervention [1,2]. In clinic, to ensure that more diagnostic information can be retrieved from PET images, physicians prefer standard-dose PET scanning which is obtained by injecting standard-dose radioactive tracers into human bodies. However, the use of radioactive tracers inevitably induces potential radiation hazards. On the other hand, reducing the tracer dose during the PET scanning will introduce unintended noise, thus leading to degraded image quality with limited diagnostic information. To tackle this clinical dilemma, it is of high interest to reconstruct standard-dose PET (SPET) images from the corresponding low-dose PET (LPET) data (i.e., sinograms or images).

In the past decade, deep learning has demonstrated its promising potential in the field of medical images [3-6]. Along the research direction of PET reconstruction, most efforts have been devoted to indirect reconstruction methods [7-16] which leverage the LPET images pre-reconstructed from the original projection data (i.e., LPET sinograms) as the starting point to estimate SPET images. For example, inspired by the preeminent performance of generative adversarial network (GAN) in computer vision [17, 18], Wang *et al.* [9] proposed a 3D conditional generative adversarial network (3D-cGAN) to convert LPET images to SPET images. However, beginning from the pre-reconstructed LPET images rather than the original LPET sinograms, these indirect methods may lose or blur details such as edges and small-size organs in the pre-reconstruction process, leading to unstable and compromised performance.

To remedy the above limitation, several studies focus on the more challenging direct reconstruction methods [19-27] which complete the reconstruction from the original sinogram domain (i.e., LPET sinograms) to the image domain (i.e., SPET images). Particularly, Haggstrom *et al.* [19] proposed DeepPET, employing a convolutional neural network (CNN)-based encoder-decoder network to reconstruct SPET images from LPET sinograms. Although these direct methods achieve excellent performance, they still have the following limitations. First, due to the lack of consideration for the boundaries, the reconstruction from the sinogram domain to the image domain often leads to distortion of the reconstructed image in the high-frequency part of the frequency domain, which is manifested as blurred edges. Second, current networks ubiquitously employ CNN-based architecture which is limited in modeling long-range semantic dependencies in data. Lacking such non-local contextual information, the reconstructed images may suffer from missing or inaccurate global structure.

In this paper, to resolve the first limitation above, we propose to represent the reconstructed SPET images in the frequency domain, then encourage them to resemble the corresponding real SPET images in the high-frequency part. As for the second limitation, we draw inspiration from the remarkable progress of vision transformer [28] in medical image analysis [29, 30]. Owing to the intrinsic self-attention mechanism, the transformer can easily correlate distant regions within the data and capture non-local information. Hence, the transformer architecture is considered in our work.

Overall, we propose an end-to-end transformer model dubbed TriDo-Former that unites triple domains of sinogram, image, and frequency to directly reconstruct the clinically acceptable SPET images from LPET sinograms. Specifically, our TriDo-Former is comprised of two cascaded transformers, i.e., a sinogram enhancement transformer (SE-Former) and a spatial-spectral reconstruction transformer (SSR-Former). The SE-Former aims to predict denoised SPET-like sinograms from LPET sinograms, so as to prevent the noise in sinograms from propagating into the image domain. Given that each row of the sinogram is essentially the projection at a certain imaging views angle, dividing it into 2D patches and feeding them directly into the transformer will inevitably break the continuity of each projection view. Therefore, to retain the inner-structure of sinograms and filter the noise, we split a sinogram by rows and obtain a set of 1D sequences of different imaging view angles. Then, the relations between view angles are modeled via the self-attention mechanism in the SE-Former. Note that the SE-Former is designed specifically for the sinogram domain of LPET to effectively reduce noise based on the imaging mechanisms of PET. The denoised sinograms can serve as a better basis for the subsequent sinogram-to-image reconstruction. The SSR-Former is designed to reconstruct SPET images from the denoised sinograms. In pursuit of better image quality, we construct the SSR-Former by adopting the powerful swin transformer [31] as the backbone. To compensate for the easily lost high-frequency details, we propose a global frequency parser (GFP) and inject it into the SSR-Former. The GFP acts as a learnable frequency filter to globally modify the components of specific frequencies of the frequency domain, forcing the network to learn accurate high-frequency details and produce construction results with shaper boundaries. Through the above triple-domain supervision, our TriDo-Former exhausts the model representation capability, thereby achieving better reconstructions.

The contributions of our proposed method can be described as follows. (1) To fully exploit the triple domains of sinogram, image, and frequency while capturing global context, we propose a novel triple-domain transformer to directly reconstruct SPET images from LPET sinograms. *To our knowledge, we are the first to leverage both triple-domain knowledge and transformer for PET reconstruction.* (2) We develop a sinogram enhancement transformer (SE-Former) that is tailored for the sinogram domain of LPET to suppress the noise while maintaining the inner-structure, thereby preventing the noise in sinograms from propagating into the image domain during the sinogram-to-image reconstruction. (3) To reconstruct high-quality PET images with clear-cut details, we design a spatial-spectral transformer (SS-Former) incorporated with the global frequency parser (GFP) which globally calibrates the frequency components in the frequency domain for recovering high-frequency details. (4) Experimental results demonstrate the superiority of our method both qualitatively and quantitatively, compared with other state-of-the-art methods.

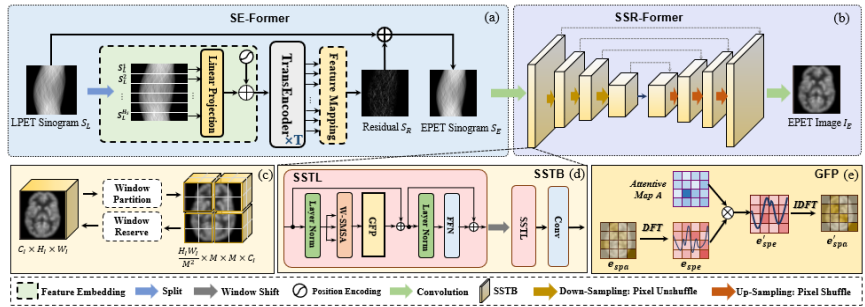

**Fig. 1.** Overview of the proposed TriDo-Former.

## 2   Methodology

The overall architecture of our proposed TriDo-Former is depicted in Fig. 1, which consists of two cascaded sub-networks, i.e., a sinogram enhancement transformer (SE-Former) and a spatial-spectral reconstruction transformer (SSR-Former). Overall, taking the LPET sinograms as input, the SE-Former first predicts the denoised SPET-like sinograms which are then sent to SSR-Former to reconstruct the estimated PET (denoted as EPET) images. A detailed description is given in the following sub-sections.

### 2.1   Sinogram Enhancement Transformer (SE-Former)

As illustrated in Fig.1 (a), the SE-Former which is responsible for denoising in the input LPET sinograms consists of three parts, i.e., a feature embedding module, transformer encoder (TransEncoder) blocks, and a feature mapping module. Given that each row of sinogram is the 1D projection at an imaging view angle, we first divide the LPET sinograms by rows and perform linear projection in the feature embedding module to obtain a set of 1D sequences, each contains consistent information of a certain view angle. Then, we perform self-attention in the TransEncoder blocks to model the interrelations between projection view angles, enabling the network to better model the general characteristics under different imaging views which is crucial for sinogram denoising. After that, the feature mapping module predicts the residual between the LPET and SPET sinograms which is finally added to the input LPET sinograms to generate the EPET sinograms as the output of SE-Former. We argue that the introduction of residual learning allows the SE-Former to focus only on learning the difference between LPET and SPET sinograms, facilitating faster convergence.

**Feature Embedding:** We denote the input LPET sinogram as $S_L \in \mathbb{R}^{C_s \times H_s \times W_s}$, where $H_s$, $W_s$ are the height, width and $C_s$ is the channel dimension. As each row of sinogram is a projection view angle, the projection at the $i$-th ($i = 1,2, \dots H_s$) row can be defined as $s_L^i \in \mathbb{R}^{C_s \times W_s}$. Therefore, by splitting the sinogram by rows, we obtain a set of 1D sequence data $S_L^* = \{s_L^i\}_{i=1}^{H_s} \in \mathbb{R}^{H_s \times D}$, where $H_s$ is the number of projection view angles and $D = C_s \times W_s$ equals to the pixel number in each sequence data. Then, $S_L^*$ is linearly projected to sequence $\tilde{S}_L^* \in \mathbb{R}^{H_s \times d}$, where $d$ is the output dimension of the projection. To maintain the position information of different view angles, we introduce a learnable position embedding $S_{pos} = \{s_{pos}^i\}_{i=1}^{H_s} \in \mathbb{R}^{H_s \times d}$ and fuse it with $\tilde{S}_L^*$ by element-wise addition, thus creating the input feature embedding $F_0 = S_{pos} + \tilde{S}_L^*$ which is further sent to $T$ stacked TransEncoder blocks to model global characteristics between view angles.

**TransEncoder:** Following the standard transformer architecture [28], each TransEncoder block contains a multi-head self-attention (MSA) module and a feed forward network (FFN) respectively accompanied by layer normalization (LN). For $j$-th ($j = 1,2,\dots,T$) TransEncoder block, the calculation process can be formulated as:

$$F_j = F_{j-1} + MSA\left(LN(F_{j-1})\right) + FFN(LN(F_{j-1} + MSA(LN(F_{j-1})))), \quad (1)$$

where $F_j$ denotes the output of $j$-th TransEncoder block. After applying $T$ identical TransEncoder blocks, the non-local relationship between projections at different view angles is accurately preserved in the output sequence $F_T \in \mathbb{R}^{H_s \times d}$.

**Feature Mapping:** The feature mapping module is designed for projecting the sequence data back to the sinogram. Concretely, $F_T$ is first reshaped to $\mathbb{R}^{C' \times H_s \times W_s}$ ($C' = \frac{d}{W_s}$) and then fed into a linear projection layer to reduce the channel dimension from $C'$ to $C_s$. Through these operations, the residual sinogram $S_R \in \mathbb{R}^{C_s \times H_s \times W_s}$ of the same dimension as $S_L$, is obtained. Finally, following the spirit of residual learning, $S_R$ is directly added to the input $S_L$ to produce the output of SE-Former, i.e., the predicted denoised sinogram $S_E \in \mathbb{R}^{C_s \times H_s \times W_s}$.

## 2.2 Spatial-Spectral Reconstruction Transformer (SSR-Former)

The SSR-Former is designed to reconstruct the denoised sinogram obtained from the SE-Former to the corresponding SPET images. As depicted in Fig.1 (b), SSR-Former adopts a 4-level U-shaped structure, where each level is formed by a spatial-spectral transformer block (SSTB). Furthermore, each SSTB contains two spatial-spectral transformer layers (SSTLs) and a convolution layer for both global and local feature extraction. Meanwhile, a $3 \times 3$ convolution is placed as a projection layer at the beginning and the end of the network. For detailed reconstruction and invertibility of sampling, we employ the pixel-unshuffle and pixel-shuffle operators for down-sampling and up-sampling. In addition, skip connections are applied for multi-level feature aggregation.

**Spatial-Spectral Transformer Layer (SSTL):** As shown in Fig. 1(d), an SSTL consists of a window-based spatial multi-head self-attention (W-SMSA) followed by FFN and LN. Following swin transformer [31], a window shift operation is conducted between the two SSTLs in each SSTB for cross-window information interactions. Moreover, to capture the high-frequency details which can be easily lost, we devise global frequency parsers (GFPs) that encourage the model to recover the high-frequency component of the frequency domain through the global adjustment of specific frequencies. Generally, the W-SMSA is leveraged to guarantee the essential global context in the reconstructed PET images, while GFP is added to enrich the high-frequency boundary details. The calculations of the core W-SMSA and GFP are described as follows.

**Window-based Spatial Multi-Head Self-Attention (W-SMSA):** Denoting the input feature embedding of certain W-SMSA as $e_{in} \in \mathbb{R}^{C_I \times H_I \times W_I}$, where $H_I$, $W_I$ and $C_I$ represent the height, width and channel dimension, respectively. As depicted in Fig. 1(c), a window partition operation is first conducted in spatial dimension with a window size of $M$. Thus, the whole input features are divided into $N$ ($N = \frac{H_I \times W_I}{M^2}$) non-overlapping patches $e_{in}^* = \{e_{in}^m\}_{m=1}^N$. Then, a regular spatial self-attention is performed separately for each window after partition. After that, the output patches are gathered through the window reverse operation to obtain the spatial representative feature $e_{spa} \in \mathbb{R}^{C_I \times H_I \times W_I}$.

**Global Frequency Parser (GFP):** After passing the W-SMSA, the feature $e_{spa}$ are already spatially representative, but still lack accurate spectral representations in the frequency domain. Hence, we propose a GFP module to rectify the high-frequency component in the frequency domain. As illustrated in Fig. 1(e), the GFP module is comprised of a 2D discrete Fourier transform (DFT), an element-wise multiplication between the frequency feature and the learnable global filter, and a 2D inverse discrete Fourier transform (IDFT). Our GFP can be regarded as a learnable version of frequency filters. The main idea is to learn a parameterized attentive map applying on the frequency domain features. Specifically, we first convert the spatial feature $e_{spa}$ to the frequency domain via 2D DFT, obtaining the spectral feature $e_{spe} = DFT(e_{spa})$. Then, we modulate the frequency components of $e_{spe}$ by multiplying a learnable parameterized attentive map $A \in \mathbb{R}^{C_I \times H_I \times W_I}$ to $e_{spe}$, which can be formulated as:

$$e'_{spe} = A \cdot e_{spe}, \qquad (2)$$

The parameterized attentive map $A$ can adaptively adjust the frequency components of the frequency domain and compel the network to restore the high-frequency part to resemble that of the supervised signal, i.e., the corresponding real SPET images (ground truth), in the training process. Finally, we reverse $e'_{spe}$ back to the image domain by adopting 2D IDFT, thus obtaining the optimized feature $e'_{spa} = DFT(e'_{spe})$. In this manner, more high-frequency details are preserved for generating shaper constructions.

### 2.3 Objective Function

The objective function for our TriDo-Former is comprised of two aspects: 1) a sinogram domain loss $L_{sino}$ and 2) an image domain loss $L_{img}$.

The sinogram domain loss aims to narrow the gap between the real SPET sinograms $S_S$ and the EPET sinograms $S_E$ that are denoised from the input LPET sinograms. Considering the critical influence of sinogram quality, we apply the L2 loss to increase the error punishment, thus forcing a more accurate prediction. It can be expressed as:

$$L_{sino} = E_{S_S,S_E \sim p_{data(S_S,S_E)}} ||S_S - S_E||_2 , \qquad (3)$$

For the image domain loss, the L1 loss is leveraged to minimize the error between the SPET images $I_S$ and the EPET images $I_E$ while encouraging less blurring, which can be defined as:

$$L_{img} = E_{I_S,I_E \sim p_{data(I_S,I_E)}} ||I_S - I_E||_1 , \qquad (4)$$

Overall, the final objective function is formulated by the weighted sum of the above losses, which is defined as:

$$L_{total} = L_{sino} + \lambda L_{img}. \qquad (5)$$

where $\lambda$ is the hyper-parameters to balance these two terms.

### 2.4 Details of Implementation

Our network is implemented by Pytorch framework and trained on an NVIDIA GeForce GTX 3090 with 24 GB memory. The whole network is trained end-to-end for 150 epochs in total using Adam optimizer with the batch size of 4. The learning rate is initialized to 4e-4 for the first 50 epochs and decays linearly to 0 for the remaining 100 epochs. The number $T$ of the TransEncoder in SE-Former is set to 2 and the window size $M$ is set to 4 in the W-SMSA of the SSR-Former. The weighting coefficient $\lambda$ in Eq. (6) is empirically set as 10.

## 3    Experiments and Results

**Datasets:** We train and validate our proposed TriDo-Former on a real human brain dataset including 8 normal control (NC) subjects and 8 mild cognitive impairment (MCI) subjects. All PET scans are acquired by a Siemens Biograph mMR system housed in Biomedical Research Imaging Center. A standard dose of 18F-Flurodeoxy-glucose ([$^{18}$F] FDG) was administered. According to standard protocol, SPET sinograms were acquired in a 12-minute period within 60-minute of radioactive tracer injection, while LPET sinograms were obtained consecutively in a 3-minute shortened acquisition time to simulate the acquisition at a quarter of the standard dose. The SPET images which are utilized as the ground truth in this study were reconstructed from the corresponding SPET sinograms using the traditional OSEM algorithm [32].

**Experimental Settings:** Due to the limited computational resources, we slice each 3D scan of size 128 × 128 × 128 into 128 2D slices with a size of 128 × 128. The Leave-One-Out Cross-Validation (LOOCV) strategy is applied to enhance the stability of the model with limited samples. To evaluate the performance, we adopt three typical quantitative evaluation metrics including peak signal-to-noise (PSNR), structural similarity index (SSIM), and normalized mean squared error (NMSE). Note that, we restack the 2D slices into complete 3D PET scans for evaluation.

**Table 1.** Quantitative comparison with five PET reconstruction methods in terms of PSNR, SSIM, and NMSE. The best performance is marked as **bold**.

| Method | NC subject | | | MCI subject | | | Params | GFLOPs |
|---|---|---|---|---|---|---|---|---|
| | PSNR | SSIM | NMSE | PSNR | SSIM | NMSE | | |
| OSEM [32] | 20.684 | 0.979 | 0.0530 | 21.541 | 0.977 | 0.0580 | - | - |
| DeepPET [19] | 23.991 | 0.982 | 0.0248 | 24.125 | 0.982 | 0.0272 | 60M | 49.20 |
| Sino-cGAN [23] | 24.191 | 0.985 | 0.0254 | 24.224 | 0.985 | 0.0269 | 39M | 19.32 |
| LCPR-Net [24] | 24.313 | 0.985 | 0.0227 | 24.607 | 0.985 | 0.0257 | 77M | 77.26 |
| 3D-cGAN [9] | 24.024 | 0.983 | 0.0231 | 24.617 | 0.981 | 0.0256 | 127M | 70.38 |
| **Proposed** | **24.912** | **0.987** | **0.0203** | **25.288** | **0.987** | **0.0228** | **38M** | **16.05** |

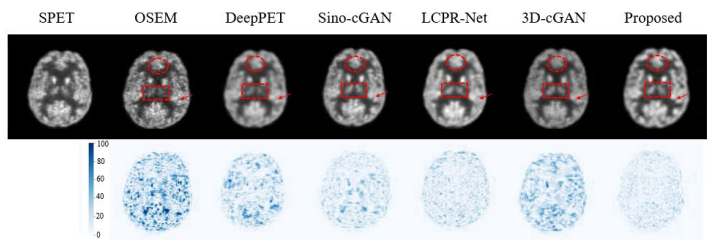

**Fig. 2.** Visual comparison of the reconstruction methods.

**Comparative Experiments:** We compare our TriDo-Former with four direct reconstruction methods, including (1) OSEM [32] (applied on the input LPET sinograms, serving as the lower bound), (2) DeepPET [19], (3) Sino-cGAN [23], and (4) LCPR-Net [24] as well as one indirect reconstruction methods, i.e., (5) 3D-cGAN [9]. The comparison results are given in Table 1, from which we can see that our TriDo-Former achieves the best results among all the evaluation criteria. Compared with the current state-of-the-art LCPR-Net, our proposed method still enhances the PSNR and SSIM by 0.599 dB and 0.002 for NC subjects, and 0.681 dB and 0.002 for MCI subjects, respectively. Moreover, our model also has minimal parameters and GLOPs of 38M and 16.05, respectively, demonstrating its speed and feasibility in clinical applications. We also visualize the results of our method and the compared approaches in Fig. 2, where the differences in global structure are highlighted with circles and boxes while the differences in edge details are marked by arrows. As can be seen, compared with other methods which have inaccurate structure and diminished edges, our TriDo-Former yields the best visual effect with minimal error in both global structure and edge details.

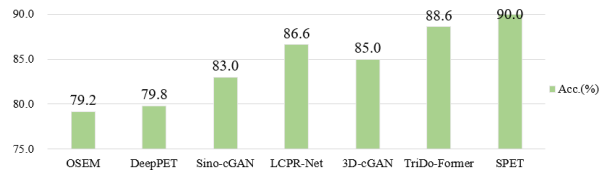

**Fig. 3.** Results of the clinical diagnosis of Alzheimer's disease (NC/MCI).

**Evaluation on clinical diagnosis:** To further prove the clinical value of our method, we further conduct an Alzheimer's disease diagnosis experiment as the downstream task. Specifically, a multi-layer CNN is firstly trained by real SPET images to distinguish between NC and MCI subjects with 90% accuracy. Then, we evaluate the PET images reconstructed by different methods on the trained classification model. Our insight is that, if the model can discriminate between NC and MCI subjects from the reconstructed images more accurately, the quality of the reconstructed images and SPET images (whose quality is preferred in clinical diagnosis) are closer. As shown in Fig. 3, the classification accuracy of our proposed method (i.e., 88.6%) is the closest to that of SPET images (i.e., 90.0%), indicating the huge clinical potential of our method in facilitating disease diagnosis.

**Table 2.** Quantitative comparison with models constructed in the ablation study in terms of PSNR, SSIM, and NMSE.

| Method | NC subjects | | | MCI subjects | | |
|---|---|---|---|---|---|---|
| | PSNR | SSIM | NMSE | PSNR | SSIM | NMSE |
| DnCNN + UNet | 23.872 | 0.981 | 0.0253 | 24.153 | 0.982 | 0.0266 |
| SE-Former + UNet | 24.177 | 0.982 | 0.0249 | 24.506 | 0.982 | 0.0257 |
| Proposed w/o GFP | 24.583 | 0.984 | 0.0235 | 24.892 | 0.984 | 0.0250 |
| **Proposed** | **24.912** | **0.987** | **0.0203** | **25.288** | **0.987** | **0.0228** |

**Ablation study:** To verify the effectiveness of the key components of our TriDo-Former, we conduct the ablation studies with the following variants: (1) replacing SE-Former and SSR-Former with DnCNN [33] (the famous CNN-based denoising network) and vanilla U-Net (denoted as DnCNN + UNet), (2) replacing DnCNN with SE-Former (denoted as SE-Former + UNet), (3) replacing the U-Net with our SSR-Former but removing GFP (denoted as Proposed w/o GFP), and (4) using the proposed TriDo-Former model (denoted as Proposed). According to the results in Table 2, the performance of our model progressively improves with the introduction of SE-Former and SSR-Former. Particularly, when we remove the GFP in SSR-Former, the performance largely decreases as the model fails to recover high-frequency details. Moreover, we conduct the clinical diagnosis experiment and the spectrum analysis to further prove the effectiveness of the GFP, and the results are included in supplementary material.

## 4     Conclusion

In this paper, we innovatively propose a triple-domain transformer, named TriDo-Former, for directly reconstructing the high-quality PET images from LPET sinograms. Our model exploits the triple domains of sinogram, image, and frequency as well as the ability of the transformer in modeling long-range interactions, thus being able to reconstruct PET images with accurate global context and sufficient high-frequency details. Experimental results on the real human brain dataset have demonstrated the feasibility and superiority of our method, compared with the state-of-the-art PET reconstruction approaches.

**Acknowledgement.** This work is supported by the National Natural Science Foundation of China (NSFC 62071314), Sichuan Science and Technology Program 2023YFG0263, 2023NSFSC0497, 22YYJCYJ0086, and Opening Foundation of Agile and Intelligent Computing Key Laboratory of Sichuan Province.